\documentclass[%
 reprint,
 amsmath,amssymb,
 aps,prl,
]{revtex4-2}

\usepackage{graphicx}
\usepackage{dcolumn}
\usepackage{bm}
\usepackage{xcolor}
\usepackage{soul}
\usepackage{epstopdf}

\begin{document}

\preprint{APS/123-QED}

\title{Tristability in viscoelastic flow past side-by-side microcylinders}

\author{Cameron C. Hopkins}
\author{Simon J. Haward}
\author{Amy Q. Shen}
\affiliation{Okinawa Institute of Science and Technology Graduate University, Onna-son, Okinawa, 904-0495, Japan
}%

\date{\today}

\begin{abstract}
Viscoelastic flows through microscale porous arrays exhibit complex path-selection and switching phenomena. However, understanding this process is limited by a lack of studies linking between a single object and large arrays. Here, we report experiments on viscoelastic flow past side-by-side microcylinders with variable intercylinder gap. With increasing flow rate, a sequence of two imperfect symmetry-breaking bifurcations forces selection of either one or two of the three possible flow paths around the cylinders. Tuning the gap length through the value where the first bifurcation becomes perfect reveals regions of bi and tristability in a dimensionless flow rate-gap length ‘phase’ diagram.
\end{abstract}

\maketitle

Since the advent of microfluidics in the early 2000s~\cite{Stone2004,Squires2005}, geometries with length-scales $\ell \sim O(100~\mu$m) have become a vital tool in experimental fluid dynamics. At the microscale, viscoelastic fluids (with properties between viscous liquids and elastic solids) can flow with negligible inertia (Reynolds number $\textnormal{Re} \sim \ell \ll 1$), but high elasticity (Weissenberg number $\textnormal{Wi} \sim \ell^{-1} \gg 1$)~\cite{Squires2005}. In such flows, elasticity becomes the dominant source of nonlinearity, leading to instabilities~\cite{Arratia2006,Burshtein2017,Dey2018,Haward2019,Haward2020,Hopkins2020,Varchanis2019}, and time-dependency that impact widespread processes ranging from jet fragmentation~\cite{Wei2015,Keshavarz2016} to hemodynamics~\cite{Brust2013,Thiebaud2014} and porous media flows~\cite{Browne2020,Walkama2020,Muller1998,Eberhard2020,De2017,Kawale2017,Ekanem2020}. In particular, the path selection and switching phenomena in viscoelastic porous media flow is considered of fundamental importance in processes such as enhanced oil recovery, groundwater remediation, filtration, and drug delivery ~\cite{Browne2020,Walkama2020,Muller1998,Eberhard2020,De2017,Kawale2017}. 

Porous media are frequently modeled by ordered and disordered arrays of microfluidic circular cylinders~\cite{Walkama2020,Muller1998,Eberhard2020,De2017,Kawale2017}. Flow past a single circular cylinder in a channel is an archetypal problem in fluid dynamics, and a `benchmark' for studying viscoelastic flows. The stagnation point downstream of a cylinder is a location where streamline curvature combines with strong velocity gradients; conditions that render viscoelastic base flows prone to instability and downstream fluctuations~\cite{Pakdel1996,McKinley1996,Pan2013,Qin2017}. For fluids with a shear-rate-dependent viscosity (i.e., shear-thinning), the perturbation to the base flow can lead to a steady symmetry-breaking flow bifurcation where the viscoelastic fluid selects a preferred path around one, or other, side of the cylinder~\cite{Dey2018,Haward2019,Haward2020,Varchanis2020}. This behavior has clear relevance to understanding transport through porous arrays, but the interaction with neighboring array elements is lacking. Building `bottom-up' complexity towards more realistic model systems, it is natural to consider two cylindrical objects either aligned in the flow direction, or positioned side-by-side in a channel. Viscoelastic flow past two (or more) objects aligned on the flow axis is a well studied problem~(e.g., Refs.~\onlinecite{Pan2013,Shi2016,Haward2018,Hopkins2020}). However, although equally important, the case of two objects positioned transverse to the flow has received scant attention, with only one numerical study conducted at high Reynolds number~\cite{Peng2020}. To date, creeping viscoelastic flow past side-by-side cylinders has not been studied. 

\begin{figure}
    \centering
    \includegraphics[width=8.6cm]{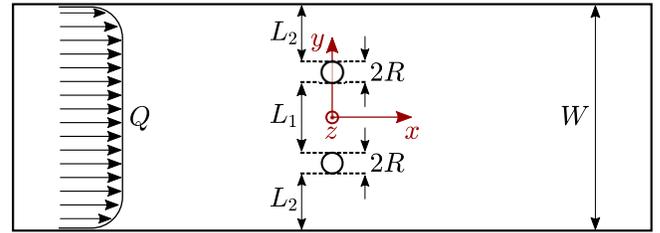}
\vspace{-0.2in}
    \caption{Schematic diagram of the $x-y$ plane of the microfluidic channels. Flow is left-to-right at volumetric rate $Q$.}
    \label{fig:schematic}
\end{figure}

In this Letter, we present microfluidic experiments of a viscoelastic shear-thinning fluid flowing past two microcylinders transverse to the primary flow direction (Fig.~\ref{fig:schematic}) and show that the resulting non-linear flow behavior at high Wi is significantly influenced by the spacing of the cylinders. We show that, due to a combination of supercritical bifurcations that occur as Wi is varied, multiple stable flow states are possible in a given geometry. This is the first study of low-Re viscoelastic flow in such a geometry and serves as a fundamental contribution towards understanding deterministic path-selection in porous media flow.

The model viscoelastic fluid is a well-studied aqueous wormlike micellar (WLM) solution consisting of 100 mM cetylpyridinium chloride (CPyCl) and 60 mM sodium salicylate (NaSal)~\cite{Rehage1988,Rehage1991}. At $24^{\circ}$C (ambient laboratory temperature), the entangled WLM solution has a zero shear viscosity $\eta_{0} = 47$~Pa~s, exhibits a stress-plateau (shear-banding region~\cite{Fielding2016}), and in small-amplitude oscillation is well-described by a single-mode Maxwell model with relaxation time $\lambda = 1.7$ s~(Fig.~S1~\cite{ESI}). 

Microfluidic channels (Fig.~\ref{fig:schematic}) were fabricated in fused silica by selective laser-induced etching~\cite{Burshtein2019}. 
The eleven channels used all have a rectangular cross-section with width $W = 400~\mu$m transverse to the flow ($y$-direction) and height $H = 2000~\mu$m in the neutral ($z$) direction. Each channel contains two cylinders of radius $R=20~\mu$m equally spaced either side of the primary flow ($x$) axis. The intercylinder separation $L_1$ is varied between channels in the range  $107 < L_1 < 147~\mu$m. The spacing between the cylinders and the channel sidewalls is $L_{2} = (W-L_1-4R)/2$, and we define a dimensionless gap ratio $G = L_{1}/(L_{1} + L_{2})$. This parameter in principle spans $0 < G < 1$, where $G = 0$ implies the two cylinders are touching at the channel centerline, while $G = 1$ implies the cylinders are touching opposite channel walls. The channels used span a range $0.50 \leq G \leq 0.62$, which encompasses the full range of flow behavior. 

\begin{figure}[t]
    \centering
    \includegraphics[width=8.6cm]{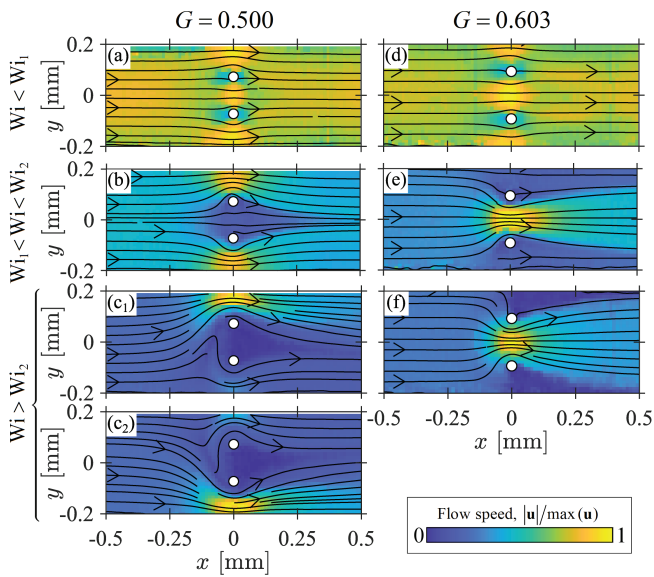}
\vspace{-0.2in}
    \caption{Evolution of velocity fields with Wi for the WLM solution in channels with contrasting values of $G = 0.500$ (a-c) and $G = 0.603$ (d-f). c$_{1}$ and c$_{2}$ indicate the two possible states for $G = 0.500$ and $\textnormal{Wi} > \textnormal{Wi}_2$.}
    \label{fig:Vfields}
\end{figure}

Flow is driven by syringe pumps (Cetoni GmbH) programmed to impose quasistatic variations in the volumetric flow rate $Q$, hence average flow velocity ${U=Q/WH}$, and Weissenberg number Wi $= \lambda U/R$. 
Quantitative spatially-resolved flow fields are obtained using micro-particle image velocimetry ($\mu$-PIV, TSI Inc., MN~\cite{Wereley2005,Wereley2010}). At each imposed Wi, the motion of a low concentration of fluorescent seeding particles ($2~\mu$m diameter) is captured at the channel half-height ($z=0$ plane) using an inverted microscope (Nikon Ti) with a $5\times$, NA = 0.15 numerical aperture objective lens and a high speed camera (Phantom Miro) working in frame-straddling mode at 25~Hz. Cross-correlation between images yields velocity vectors ${\bf{u}} = (u,v)$. Since the flows examined are all time invariant, data are ensemble averaged over a 6~s sampling window. Due to shear-localization at the channel walls, the flow profile is essentially plug-like over most of the channel cross-section~\cite{Haward2019}. Therefore, the shear-rate near the cylinders is small and we define $\textnormal{Re} = \rho U R/\eta_{0}$, where $\rho = 1000$~kg~m$^{-3}$ is the fluid density. In all experiments, $\textnormal{Re} \lesssim 10^{-4}$.

Flow fields representative of those observed as Wi is varied are shown in Fig.~\ref{fig:Vfields} using two channels with contrasting $G$. Fig.~\ref{fig:Vfields}a-c and Fig.~\ref{fig:Vfields}d-f illustrate the behavior for `small' and `large' $G$, respectively.  Irrespective of $G$, for low $\textnormal{Wi} < \textnormal{Wi}_1 \approx 15$ (Figs.~\ref{fig:schematic}a,d), elastic and inertial forces are small and the flow is dominated by the viscous force. Flow is approximately symmetric about $x = 0$ and $y = 0$, and fluid passes through all three available gaps. For `small' $G = 0.500$, as $\textnormal{Wi}$ exceeds $\textnormal{Wi}_{1}$ (Fig.~\ref{fig:Vfields}b), elasticity dominates and the system undergoes a first transition from the low-Wi symmetric state to a diverging `D' state where the fluid avoids the gap between the cylinders and flows symmetrically around their sides. The velocity field is qualitatively similar to that for viscoelastic flow around a single obstacle~\cite{Haward2018,Haward2019}. For a Newtonian fluid, two objects appear as one when the ratio of separation to radius, $L_{1}/R < 0.2$~\cite{Supradeepan2014}, whereas here the ratio is much greater at $L_1/R >5$. Further increasing Wi, the system undergoes a second transition at $\textnormal{Wi}_2 \approx 50$ to an asymmetric-diverging `AD' state in which the fluid selects a single preferred path either above ($y>0$, Fig.~\ref{fig:Vfields}c$_{1}$), or below ($y<0$, Fig.~\ref{fig:Vfields}c$_{2}$) the pair of cylinders. This randomly chosen bias is also similar to that observed for viscoelastic shear-thinning fluids flowing around a single cylinder~\cite{Haward2019,Haward2020,Varchanis2020}. 

For `large' $G = 0.603$, the first transition at $\textnormal{Wi}>\textnormal{Wi}_1$ results in a converging `C' flow state where the fluid flows preferentially between the cylinders, avoiding the gaps at their sides (Fig.~\ref{fig:Vfields}e). In contrast to the small-$G$ case, as Wi increases there is no second transition at $\textnormal{Wi}_{2}$ and the C state is maintained until the flow eventually becomes time-dependent at $\textnormal{Wi}\gg\textnormal{Wi}_1$. The nature of the time-dependence is qualitatively similar to that previously reported for a single cylinder~\cite{Haward2019} and will not be discussed further here. We note that results for small and large $G$ in Fig.~\ref{fig:Vfields} using a shear-thinning, but non-shear-banding viscoelastic polymer solution, show analogous flow behavior see Figs. S2 and S3~\cite{ESI}. 

\begin{figure*}[ht]
    \centering
    \includegraphics[width=17.2cm]{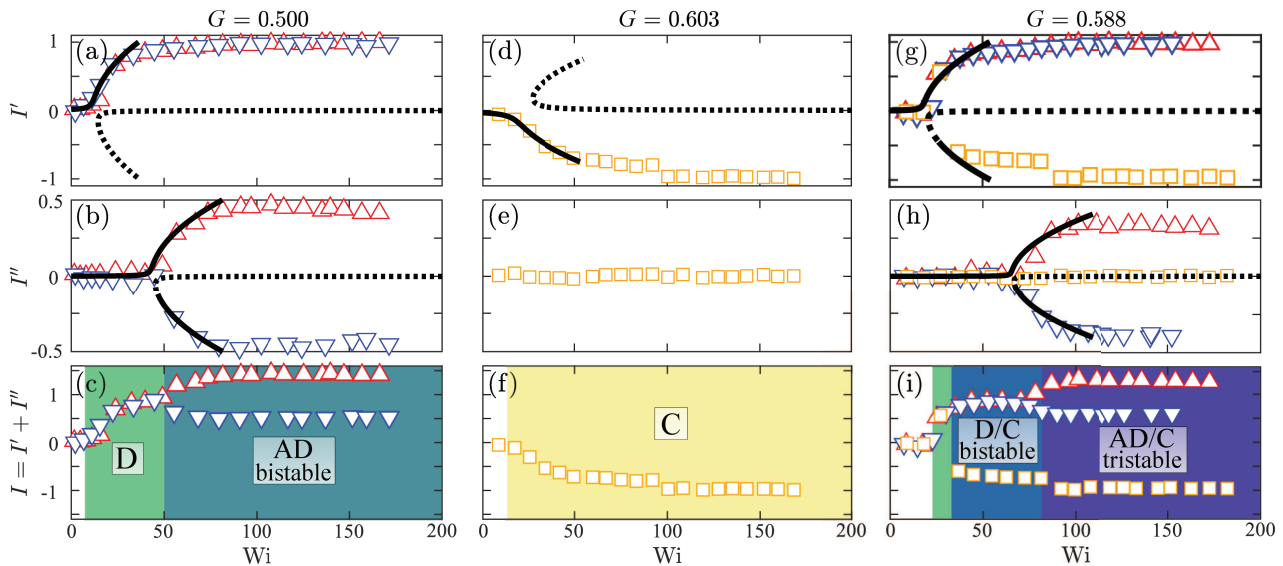}
\vspace{-0.1in}
    \caption{Flow asymmetry parameters $I'$, $I''$, and $I =  I' + I''$ \emph{vs} Wi for microfluidic channels with (a-c) $G = 0.500$, (d-f) $G = 0.603$, and (g-i) $G = 0.588$. Symbols indicate the qualitative flow states observed at high Wi for a given experiment. ($\bigtriangleup$, $\bigtriangledown$): AD state with $I''>0$ or $I''<0$, respectively, and ($\square$): C state. The solid and dotted black curves are fits of a 4th-order Landau potential to the data (see main text). Colored backgrounds in (c),(f), and (i) delineate the various flow states.}
    \label{fig:asyms}
\end{figure*}

We quantify the critical flow behavior using two dimensionless flow asymmetry parameters $I'$ and $I''$:
\begin{equation}
I' = \frac{\frac{1}{2}(\bar u_{+}+\bar u_{-})- \bar u_{0}}{\frac{1}{2}(\bar u_{+}+\bar u_{-})+\bar u_{0}} \quad\text{and}\quad I'' = \frac{\bar u_{+}-\bar u_{-}}{\bar u_{+}+\bar u_{-}+\bar u_{0}}.
\end{equation}
Here, $\bar u_{+}$, $\bar u_{-}$, and $\bar u_{0}$ are the average values of $u$ in the upper, lower, and intercylinder gaps, respectively (see Fig.~\ref{fig:schematic}). $I'$ serves as the order parameter to quantify the first transition from the low-Wi symmetric state to either the D or C states (Fig.~\ref{fig:Vfields}b,e). $I' = 0$ when the average flow through the upper and lower gaps equals the flow through the center. Transition to the D state results in $I'>0$, since $\bar u_{0}$ decreases. Transition to the C state results in $I' < 0$, since $\bar u_{0}$ increases. $I''$ serves as the order parameter to quantify the second transition between the D and the AD states. $I'' = 0$ in the D state, since $\bar u_{+} = \bar u_{-}$ (Fig.~\ref{fig:Vfields}b). In the AD state, fluid flows preferentially through either the upper or lower gap (Fig.~\ref{fig:Vfields}c$_{1}$,c$_{2}$), resulting in $I''>0$ or $I''<0$, respectively.

The asymmetry parameters $I'$, $I''$ are shown \emph{vs} Wi in Fig.~\ref{fig:asyms} for various values of $G$ and are fitted with a quartic (double-well) Landau-type potential minimized as:

\begin{equation}\label{eq:Landau}
    \textnormal{Wi} = \textnormal{Wi}_{c}(g\epsilon^{2} + h\epsilon^{-1} + 1),
\end{equation}
where $\textnormal{Wi}_c=\textnormal{Wi}_1$ or $\textnormal{Wi}_2 $ is the critical Weissenberg number for the bifurcation, and the order parameter $\epsilon$ can be either $I'$ or $I''$. In all the fits, the growth rate coefficient $g$ is order unity, and the asymmetric term in $h$ quantifies system imperfections that bias a transition to a favored branch. The phenomenological Landau model for equilibrium phase transitions has long been found to provide a good description of bifurcation phenomena in driven nonequilibrium systems including Newtonian and viscoelastic flows~\cite{Aitta1985,Gollub1976,Burshtein2017,Haward2019}. Eq.~\ref{eq:Landau} describes forward (supercritical) pitchfork bifurcations without hysteresis.

For small $G= 0.500$, the first transition in $I'$ (Fig.~\ref{fig:asyms}a) occurs at $\textnormal{Wi}_{1} \approx 13$, and is a slightly imperfect ($h \approx -0.016$) supercritical pitchfork bifurcation where the favored branch ($I'>0$) gives diverging (D) flow. The unfavored ($I'<0$) branch was never observed, but its hypothetical existence is indicated in Fig.~\ref{fig:asyms}a by the dotted line. With increasing Wi, $I'\rightarrow1$, implying that almost no fluid passes between the two cylinders (as qualitatively evident from Fig.~\ref{fig:Vfields}b). The second transition in $I''$ (Fig.~\ref{fig:asyms}b) from the D state to the asymmetric diverging (AD) state, occurs at $\textnormal{Wi}_{2} \approx$ 44. The imperfection in this second bifurcation is very small ($h \approx -0.0013$). The favored branch gives $I'' > 0$, but the unfavored $I''<0$ branch can also be reached and followed by initiating the flow at a high Wi and subsequent quasistatic reduction. The complete bifurcation diagram showing the total asymmetry $I = I' + I''$ \emph{vs} Wi for $G = 0.500$ is shown in Fig.~\ref{fig:asyms}c. The first bifurcation results in $I \rightarrow 1$. The second bifurcation splits $I$ into two branches, $I \rightarrow 1.5$ or $I \rightarrow 0.5$.

The behavior for $G = 0.603$ is shown in Fig.~\ref{fig:asyms}d-f. In this case, the first bifurcation occurs at $\textnormal{Wi}_{1} \approx 20$ (Fig.~\ref{fig:asyms}d). The asymmetric term in Eq.~\ref{eq:Landau} is positive ($h \approx 0.033$), resulting in a preferred transition from symmetric to converging (C) flow. With increasing Wi, $I' \rightarrow -1$, indicating that nearly all of the fluid passes between the cylinders (see Fig.~\ref{fig:Vfields}e,f). Since $h$ is relatively large, the negative $I'$ branch is strongly preferred. The positive $I'$ branch (dotted line in Fig.~\ref{fig:asyms}d) is never observed experimentally. When the system selects the C state at the first transition, a second bifurcation is not observed, and $I'' \approx 0$ for all Wi (Fig.~\ref{fig:asyms}e). The complete bifurcation diagram for $G = 0.603$ is shown in Fig.~\ref{fig:asyms}f: since $I'' \approx 0$, $I \approx I'$. 

The data shown in Figs.~\ref{fig:Vfields} and~\ref{fig:asyms}a-f demonstrate two disparate flow behaviors that are sensitive to the value of $G$. The first bifurcation to either the D or C states is well described as a supercritical pitchfork bifurcation quantified by $I'$. The two states are different branches of the same bifurcation and the value of $G$ determines which branch is selected by changing the sign of the asymmetric term ($h$) in Eq.~\ref{eq:Landau}. This implies the existence of a specific intermediate value of $G$ at which the bifurcation of $I'$ should be perfect ($h=0$) and the D or C states are equally likely.

By examining a range of intermediate values $0.56 < G < 0.60$, we confirmed this assumption, as exemplified by Fig.~\ref{fig:asyms}g-i for $G = 0.588$. Here, increasing Wi quasistatically from 0, the favored positive $I'$ branch (D state) is observed (Fig.~\ref{fig:asyms}g, triangles) on exceeding $\textnormal{Wi}_{1} \approx 20$. However, the imperfection is sufficiently small ($h \approx -0.007$), that by quasistatic reduction of Wi from a high value, the unfavored C state branch (squares) can also be followed. From the D state, on exceeding $\textnormal{Wi}_{2} \approx 66$, the second bifurcation to the AD state occurs to either positive or negative $I''$ with almost equal likelihood in a given experiment (Fig.~\ref{fig:asyms}h triangles) since $h\approx 0$. The complete bifurcation diagram for $G = 0.588$ in Fig.~\ref{fig:asyms}i shows the bistable coexistence of the C and D states for $\textnormal{Wi}_1< \textnormal{Wi} <\textnormal{Wi}_{2}$. For $\textnormal{Wi} >\textnormal{Wi}_{2}$, the system is tristable, where the two AD branches and the C branch coexist.  

\begin{figure}[!t]
    \centering
    \includegraphics[width=8.6cm]{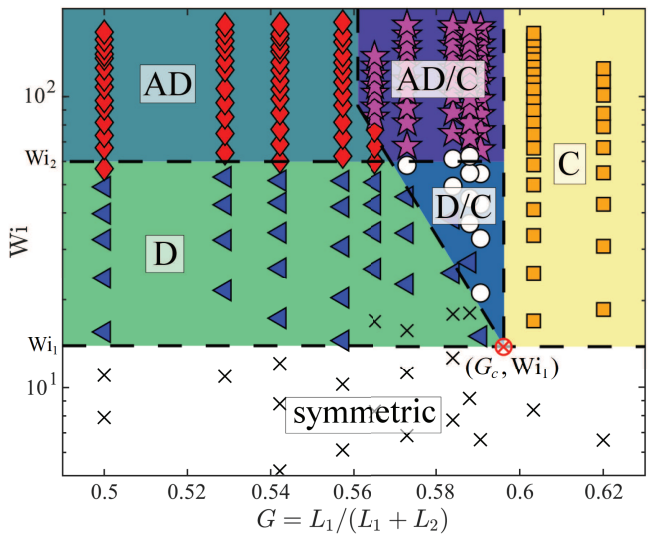}
\vspace{-0.3in}
    \caption{Flow stability diagram in Wi--$G$ state space. The dashed lines and colored shades delineate between flow states.}
    \label{fig:phasediagram}
\end{figure}

The behavior observed in all eleven microchannels is summarized in a flow stability diagram in Wi--$G$ state space (Fig.~\ref{fig:phasediagram}). For channels with `small' $G \lesssim 0.560$ the systems behave as exemplified by Figs.~\ref{fig:Vfields}a-c and~\ref{fig:asyms}a-c, with a first bifurcation at $\textnormal{Wi}_{1}$ that is biased to the D state and a second bifurcation at $\textnormal{Wi}_{2}$ to the bistable AD state with either positive or negative $I''$. For `large' $G \gtrsim 0.595$ the systems behave as shown by Figs.~\ref{fig:Vfields}d-f and~\ref{fig:asyms}d-f; the bifurcation at $\textnormal{Wi}_{1}$ is biased to the C state (which is maintained until the onset of time-dependence at $\textnormal{Wi}>>\textnormal{Wi}_{1}$). For a narrow range of $0.560 \lesssim G \lesssim 0.595$, beyond $\textnormal{Wi}_{1}$ it is possible to pass through a region of bistability between the D and C states before entering a tristable region beyond $\textnormal{Wi}_{2}$ comprising the bistable AD state and the C state. It is assumed that the D/C bistable region meets the low-Wi symmetric region at a hypothetical single point ($G_c,\textnormal{Wi}_{1}$) in the state space (red crossed circle, Fig.~\ref{fig:phasediagram}) where the first bifurcation would be perfect ($h=0$). To the left of this point, the diagonal boundary between the D/C and D regions reflects the increasing imperfection of the first bifurcation with decreasing $G$. However, based on our experiments, to the right of $G_c$ the boundary between the D/C and the C states appears to be extremely abrupt. A small increase in $G>G_{c}$ causes a significant bias to the C state in the first bifurcation.  A special case arises for $G = 0.565$, where the tristable AD/C region is reached by passing through the bistable AD region, avoiding the bistable D/C region. Apparently, at this value of $G$ and Wi, the selected flow path can spontaneously switch from an edge to the center gap, and \textit{vice-versa} for decreasing Wi.

Our extensive microfluidic experiments reveal the complex dynamical behavior of viscoelastic flow past side-by-side microcylinders with variable spacing. We observe flow transitions that are well-described by the Landau model as supercritical pitchfork bifurcations.  If the intercylinder spacing is small, a first bifurcation is biased to favour a diverging flow state where the fluid avoids the intercylinder gap. From here, a second bifurcation leads to an asymmetric-divergent state where flow through just one of the two side gaps is preferred. The overall behavior is reminiscent of shear-thinning viscoelastic flow around a single obstacle~\cite{Haward2019,Haward2020}. For large intercylinder spacing, the system undergoes a single bifurcation that is biased to a converging flow state where all of the fluid passes between the cylinders. For a small range of intermediate intercylinder gaps, it is possible to identify a region of bistability where the converging and diverging states coexist, which becomes a tristable region when the diverging state undergoes the second bifurcation. 

This work is the first study of the creeping viscoelastic flow past side-by-side cylinders. Despite the modest increase in geometrical complexity from flow around a single cylinder, the dynamical behavior is significantly richer and provides an important and neglected stepping stone towards understanding viscoelastic flows through more complex arrays of objects. Our results suggest a new interpretation for how viscoelastic fluids select preferred flow paths through ordered and disordered porous arrays (e.g., Ref.~\onlinecite{Walkama2020}), indicating that each individual obstacle should be considered as a bifurcation point, which can be perfect (ordered) or imperfect (disordered), depending on the spacing between array elements. 

We gratefully acknowledge the support of the Okinawa Institute of Science and Technology Graduate University (OIST) with subsidy funding from the Cabinet Office, Government of Japan, and also funding from the Japan Society for the Promotion of Science (JSPS, Grant Nos. 18K03958, 18H01135 and 20K14656) and the Joint Research Projects (JRPs) supported by the JSPS and the Swiss National Science Foundation (SNSF). 

\providecommand{\noopsort}[1]{}\providecommand{\singleletter}[1]{#1}%

\end{document}